\documentstyle[11pt,epsfig]{article}
\newcommand{\be}{\begin{equation}}
\newcommand{\ee}{\end{equation}}
\newcommand{\ba}{\begin{eqnarray}}
\newcommand{\ea}{\end{eqnarray}}

[1999/03/29 PSNFSS v.7.2
Times font as default roman
: S Rahtz]

\begin{document}

\begin{center}
\centering{\LARGE Algorithm for Model Validation: Theory and Applications}
\end{center}

\vskip .5cm
\begin{center}
\centering{\large D. Sornette$^{1,2,3}$, A. B. Davis$^4$, K. Ide$^{1,5}$,
K. R. Vixie$^6$, V. Pisarenko$^{7}$, and J. R. Kamm$^8$}
\end{center}

\vskip .5cm
\begin{center}
\centering{$^1$ Institute of Geophysics and Planetary Physics,
    University of California, Los Angeles, CA 90095, USA}
\end{center}
\begin{center}
\centering{$^2$ Department of Earth and Space Sciences,
University of California, Los Angeles, CA 90095, USA\\
and Laboratoire de Physique de la Mati\`ere Condens\'ee, CNRS UMR
6622,\\ Universit\'e de Nice-Sophia Antipolis, 06108 Nice Cedex 2,
France}
\end{center}
\begin{center}
\centering{$^3$ now at D-MTEC, ETH Zurich, CH-8032 Zurich, Switzerland}
\end{center}
\begin{center}
\centering{$^4$ Los Alamos National Laboratory,
Space and Remote Sensing Group (ISR-2),\\
Los Alamos, NM 87545, USA}
\end{center}
\begin{center}
\centering{$^5$ Department of Atmospheric and Oceanic Sciences,
University of California, Los Angeles, CA 90095, USA}
\end{center}
\begin{center}
\centering{$^6$ Los Alamos National Laboratory,
Mathematical Modeling and Analysis Group (T-7), \\
Los Alamos, New Mexico 87545, USA}
\end{center}
\begin{center}
\centering{$^7$ International Institute of Earthquake Prediction Theory and
Mathematical Geophysics, \\
Russian Academy of Sciences, Warshavskoye sh., 79, kor. 2, Moscow
113556, Russia}
\end{center}
\begin{center}
\centering{$^8$ Los Alamos National Laboratory,
Applied Science and Methods Development Group (X-1),\\
Los Alamos, New Mexico 87545, USA}
\end{center}

\vskip 1cm
\noindent
{\bf Abstract}:
Validation is often defined as the process of determining the degree to
which a model is an accurate representation of the real world from the
perspective of its intended uses. Validation is crucial as industries
and governments depend increasingly on predictions by computer models to
justify their decisions. We propose to formulate the validation of a
given model as an iterative construction process that mimics the
often implicit process occurring in the minds of scientists. We offer a
formal representation of the progressive build-up of trust in the model.
We thus replace static claims on the impossibility of validating a given
model by a dynamic process of constructive approximation. This
approach is better adapted to the fuzzy, coarse-grained nature of
validation. Our procedure factors in the degree of redundancy versus
novelty of the experiments used for validation as well as the degree to
which the model predicts the observations. We illustrate the new
methodology first with the maturation of Quantum Mechanics as the
arguably best established physics theory and then with several concrete
examples drawn from some of our primary scientific interests: a cellular
automaton model for earthquakes, a multifractal random walk model for
financial time series, an anomalous diffusion model for solar radiation
transport in the cloudy atmosphere, and a computational fluid dynamics
code for the Richtmyer-Meshkov instability.

\pagebreak
%\vskip 0.5cm
\noindent
{\bf Introduction: Model Construction and Validation}

\noindent
At the heart of the scientific endeavor, model building
involves a slow
and arduous selection process, which can be roughly represented as
proceeding according to the following steps: (1) start from observations
and/or experiments; (2) classify them according to regularities that
they may exhibit: the presence of patterns, of some order, also sometimes referred to as 
structures or symmetries, is begging
for ``explanations'' and is thus the nucleation point of modeling; (3) use inductive reasoning,
intuition, analogies, and so on, to build hypotheses from which a model \cite{modeldef} is
constructed; (4) test the model obtained in step 3
with available observations, and then
extract predictions that are tested against new observations or by
developing dedicated experiments. The model is then rejected or refined
by an iterative process, a loop going from (1) to (4). A given model is
progressively validated by the accumulated confirmations of its predictions
by repeated experimental and/or observational tests. 

Using a model requires a language, i.e., a vocabulary and syntax, to express it.
The language can be English or French to obtain predicates specifying the properties of
and/or relation with the subject(s). It can be mathematics which is arguably
the best language to formalize the relation between quantities,
structures, space and change. It can be a computer language to implement
a set of relations and instructions logically linked in a computer code to obtain
quantative outputs in the form of string of numbers.
In this later version, validation must be distinguished from verification:
whereas {\it verification}
deals with whether the simulation code correctly solves the model equations,
{\it validation} carries an additional
degree of trust in the value of the model vis-\`a-vis experiment and,
therefore, may convince one to use its predictions to explore beyond known
territories \cite{Roache}.

The validation of models is becoming a major issue as humans are
increasingly faced with decisions involving complex tradeoffs in
problems with large uncertainties, as for instance in attempts to
control the growing anthropogenic burden on the planet \cite{WWF} within
a risk-cost framework \cite{riskcost} based on predictions of models.
For policy decisions, federal, state, and local governments increasingly
depend on computer models that are scrutinized by scientific agencies to
attest to their legitimacy and reliability.  Cognizance of this trend
and its scientific implications is not lost on the engineering
\cite{Babuska} and physics \cite{Post} communities.

How does one validate a model when it makes predictions on objects that
are not fully replicated in the laboratory, either in the range of
variables, of parameters or of scales? Indeed, a potentially
far-reaching consequence of validation is to give the ``green light''
for extrapolating a body of knowledge, which is firmly established only
in some limited ranges of variables, parameters and scales. Predictive
capability is what enables us to go beyond this clearly defined domain
into a more fuzzy area of unknown conditions and outcomes. This problem
has repeatedly appeared in different guises in practically all
scientific fields. A notable domain of application is risk
assessment: see for instance the classic paper on risks \cite{KaplanGarrick},
and the instructive history of quantitative risk analysis in US regulatory practice
\cite{Rechard}, especially in the U.S. nuclear power industry 
\cite{Keeney,Helton1,Helton2,Helton3}. 

An accute 
question in risk assessment deals with the question of
quantifying the potential for a catastrophic event
(earthquake, tornado, hurricane, flood, huge solar mass ejection,
large bolide, industrial plant explosion, ecological disaster,
financial crash, economic collapse, etc.) of amplitude never yet sampled
from the knowledge of past history and present understanding. This is
crucial, for example, in the problem of scaling the physics of
material and rock rupture tested in the laboratory to the scale of earthquakes. This
is necessary for scaling the knowledge of hydrodynamical processes
quantified in the laboratory to the length and time scales relevant to
the atmospheric/oceanic weather and climate, not to mention
astrophysical systems.
%This problem is also critical in the stewardship of the nuclear arsenal
%in the absence of testing, where the problem is to go from physics-based
%models putatively validated at small-scales in the laboratory to the
%full-scale explosion of an aging nuclear weapon.
Perhaps surprisingly, the same problem arises in the evaluation of
electronic circuits \cite{Hefner}: ``The problem is that there is no
systematic way to determine the range of applicability of the models
provided within circuit simulator component libraries.''
The example of validation of electronic circuits is particularly interesting
because it identifies the origin of the difficulties inherent in validation:
the fact that the dynamics are strongly nonlinear and complex with threshold effects,
that it does not allow for a simple-minded analytic approach consisting
in testing a circuit component by component. This same
difficulty is found in validating general circulation models of the Earth's
climate or end-to-end computer simulations of complex engineering
systems such as an aircraft or a nuclear weapon. The problem is
fundamentally due to its systemic nature. 

The theory of systems, sometimes
referred to as the theory of complex systems, is characterized by the
occurrence of surprises. The biggest one may be the
phenomenon of ``emergence'' in which qualitatively new processes
or structures appear in the collective behavior of the system, while
they can rarely be derived or guessed from the behavior of each element.
The phenomenon of ``emergence'' is similar to the philosophical law on the
``transfer of the quantity into the quality.'' A full control of the
validation process requires to account for this
emergence phenomenon, because it may contribute to the epistemic
uncertainty (the uncertainty attributable to incomplete knowledge about
a phenomenon that affects our ability to model it) associated with so-called
``unknown unknowns.''

\vskip 0.5cm
\noindent
{\bf Impossibility Statements}

\noindent
For these reasons, the possibility to validate numerical models of
natural phenomena, often endorsed either implicitly or identified as
reachable goals by natural scientists in their daily work, has been
challenged; quoting Oreskes et al. \cite{Oreskes}: ``Verification and
validation of numerical models of natural systems is impossible. This is
because natural systems are never closed and because model results are
always non-unique.'' According to this view, the impossibility of
``verifying'' or ``validating'' models is not limited  to computer
models and codes but to all theories that rely necessarily on
imperfectly measured data and auxiliary hypotheses, as Sterman et al.
\cite{auxi} put it: ``Any theory is underdetermined and thus
unverifiable, whether it is embodied in a large-scale computer model or
consists of the simplest equations.'' Accordingly, many uncertainties
undermine the predictive reliability of any model of a complex natural
system in advance of its actual use. 

Such ``impossibility'' statements are reminiscent of other
``impossibility theorems.'' Consider the mathematics of algorithmic
complexity \cite{algocompl}, which provides one approach to the
study of complex systems. Following reasoning related to that
underpinning G\"odel's incompleteness theorem, most complex systems have
been proved to be computationally irreducible, i.e., the only way to
predict their evolution is to actually let them evolve in time.
Accordingly, the future time evolution of most complex systems appears
inherently unpredictable. Such sweeping statements turn out to
have basically no practical value. This is because,
in physics and other related sciences, one aims at predicting
{\it coarse-grained} properties. Only by ignoring most of molecular
detail, for example, did researchers ever develop the laws of
thermodynamics, fluid dynamics and chemistry. Physics works and is not
hampered by computational irreducibility because we only ask for approximate
answers at some coarse-grained level \cite{Buchanan}.
By developing exact coarse-grained procedures
on computationally irreducible cellular automata,
Israeli and Goldenfeld \cite{IsraGold} have demonstrated
that prediction may simply depend on finding the right level for
describing the system. More generally, we argue that 
only coarse-grained scales are of interest in practice but their
description requires effective laws which are in
general based on finer scales. In other words,
real understanding must be rooted in the ability to predict coarser
scales from finer scales, i.e., a real understanding solves the
universal micro-macro challenge. 
Similarly, we propose that validation is possible, to some degree, as
explained below.

\vskip 0.5cm
\noindent
{\bf Validation and Hypothesis Testing}

\noindent
We start by recognizing that validation is closely related to
hypothesis testing and statistical significance tests of
mathematical statistics \cite{Borovkov}, a point made previously
by several others authors  \cite{Coleman,Hills1,Hills2,Easterling-Berger,Easterling}. 
In hypothesis testing,
a null $H_0$ is compared with an alternative hypothesis $H_1$,
in their ability to explain and fit data. The result of the test is
either to  ``reject $H_0$ in favor of $H_1$'' or ``not reject $H_0$.''
One never concludes ``reject $H_1$,'' or even ``accept $H_0$ or
$H_1$.'' If one concludes ``do
not reject $H_0$,'' this does not necessarily mean that the null hypothesis
is true, it only suggests that there is not sufficient evidence against
$H_0$ in favor of $H_1$; rejecting the null hypothesis may suggest
but does not prove that the
alternative hypothesis is true, only that it is better given the data.
Thus, one can never prove
that an hypothesis is true, only that it is less effective in 
explaining the data than another
hypothesis. One can also conclude
that an hypothesis $H_1$ is not necessary and the other more parsimonious
hypothesis $H_0$ should be
favored. The alternative hypothesis $H_1$ is not rejected,
strictly speaking, but can be found unnecessary or redundant
with respect to $H_0$. This is the situation when
there are two (or several) alternative hypotheses $H_0$ and $H_1$, which
can be composite, nested, or non-nested (the technical difficulties of
hypothesis testing depends on these structures of the competing hypotheses
\cite{Gourieroux}). This illuminates the status of code comparison in
verification and validation \cite{Trucano}. Viewed in this way, it is
clear why code comparison alone, i.e., independent of comparison to
observations/experiments, is not sufficient for validation since
validation requires comparison with experiments and several other steps
described below. The analogy with hypothesis testing makes clear that
code comparison allows the selection of one code among several codes but
does not help to conclude about the validity of a given code or model
when considered as a unique entity independently of other codes or
models. We should stress that the Sandia report \cite{Trucano}
presents an even more negative view of code comparisons because 
it addresses the common practice in the computer community
that turns to code comparisons rather than real verification or
validation, without any independent referents. Here, using the 
analogy with hypothesis testing, we have taken a more positive
view of ``Code 1 versus  referent compared with Code 2 versus reference,''
leading to an inference about which code is better based on the
comparative performance with the data. While some will consider this
as real validation, this procedure does not address the challenges
raised earlier, which justifies the 
algorithm delineated in following sections.

In the theory of hypothesis testing, there is a second class of tests, called
``tests of significance,''
in which one considers a unique hypothesis $H_0$ (model), and the
alternative is ``all the
rest,'' i.e., all hypotheses that differ from $H_0$. In
that case, the conclusion of a test can be the following: ``this data sample
does not contradict the hypothesis $H_0$,'' which is of course not the same as
``the hypothesis $H_0$ is true.'' In other words, a test of
significance cannot ``accept''
an hypothesis, it
can only fail to reject it because the hypothesis is found sufficient
at some confidence level for explaining the available data. Multiplying
the tests will not help in accepting $H_0$.

Since validation must at least
contain hypothesis testing, this shows that
statements like ``verification
and validation of numerical models of natural systems is impossible''
\cite{Oreskes}
are best rephrased in the language of mathematical statistics
\cite{Borovkov}: the theory of statistical hypothesis testing has
taught mathematical and applied statisticians for decades that one can never
prove an hypothesis or a model to be true. One can only develop an
increasing trust in it by subjecting it to more and more tests
which ``do not reject it.'' We attempt to formalize below how such trust
can be built up to lead to validation viewed as an evolving process.

\vskip 0.5cm
\noindent
{\bf Validation as a Constructive Iterative Process}

\noindent
In a standard exercise of model validation, one performs an
experiment and, in parallel,
runs the calculations with the available model. Then, a comparison between
the measurements of the experiment and the outputs of the model calculations
is performed. This comparison uses some metrics controlled by
experimental feasibility, i.e., what can actually be measured. One 
then iterates by refining the model
until (admittedly subjective) satisfactory agreement is obtained.
Then, another set of measurements
is performed, which is compared with the corresponding predictions of
the model.
If the agreement is still satisfactory without modifying the model, this is
considered
progress in the validation of the model. Iterating with experiments
testing different features of the model corresponds to mimicking the
process of construction of a theory in physics \cite{Dyson}.
As the model is exposed to increasing
scrutiny and testing, the testers develop a better understanding of
the reliability (and limitations) of the model in predicting the outcome
of new experimental and/or observational set-ups. This implies that
``validation activity should be organized like a project, with goals
and requirements, a plan, resources, a schedule, and a documented 
record'' \cite{Post}.

Extending previous proposals \cite{Coleman,Hills1,Hills2,Easterling-Berger,Easterling},
we thus propose to formulate the validation problem of a given model as
an iterative construction that embodies the often implicit process
occurring in the minds of scientists:
\begin{enumerate}
\item
One starts with an a priori trust quantified by
the value $V_{\rm prior}$ in the potential value of the model.
This quantity captures the accumulated evidence thus far.
If the model is new or the validation process is just starting, take
$V_{\rm prior} = 1$.  As we will soon see, the absolute value of
$V_{\rm prior}$ is unimportant but its relative change is important.
\item
An experiment is performed, the model is set-up to calculate
what should be the outcome of the experiment, and the comparison between
these predictions and the actual measurements is made either in model
space or in observation space.
The comparison requires a choice of metrics.
\item
Ideally, the quality of the comparison between predictions and observations
is formulated as a statistical test of significance in which an hypothesis
(the model) is tested against the alternative, which is ``all the rest.''
Then, the formulation of the comparison will be either ``the model
is rejected'' (it is not compatible with the data)
or ``the model is compatible with the data.''
In order to implement this statistical test, one needs to attribute a
likelihood $p(M | y_{\rm obs})$ or, more generally, a metric-based ``grade''
that quantifies the quality
of the comparison between the predictions of the model $M$ and 
observations $y_{\rm obs}$.
This grade is compared with the reference likelihood $q$ of ``all the rest.''
Examples of implementations include the sign test
and the tolerance interval methods \cite{Pal}.
In many cases, one does not have the luxury of a likelihood;
one has then to resort to more empirical notations of how well
the model explains crucial observations. In the most complex cases, these
notations can be binary (accepted or rejected).
\item
The posterior value of the model is
obtained according to a formula of the type 
\be
V_{\rm posterior}/V_{\rm prior}  =
F\left[ p(M | y_{\rm obs}), q; c_{\rm novel} \right]~.
\label{mgler}
\ee
In this expression, $V_{\rm posterior}$ is the posterior potential, or
coefficient, of trust in the value of the model after the
comparison between the prediction of the model and the new observations
have been performed. By the action of $F(\cdots)$, $V_{\rm posterior}$
can be either larger or smaller than $V_{\rm prior}$:
in the former case, the experimental test has
increased our trust in the validity of the model;
in the later case, the experimental test has signaled
problems with the model.
One could call $V_{\rm prior}$ and $V_{\rm posterior}$ the evolving ``potential
value of our trust'' in the model or, loosely paraphrasing the theory of
decision making in economics, the ``utility'' of the model
\cite{Morgen}.
\end{enumerate}
\noindent
The transformation from the potential value
$V_{\rm prior}$ of the model before the experimental test to
$V_{\rm posterior}$ after the test
is embodied into the multiplier $F$, which can be either larger than
$1$ (towards validation) or smaller than $1$ (towards invalidation).
We postulate that $F$ depends
on the grade $p(M | y_{\rm obs})$,
to be interpreted as proportional to the probability of the model $M$ given
the data $y_{\rm obs}$.  It is natural to compare this
probability with the reference likelihood $q$ that one or more of all
other conceivable models is compatible with the same data.

The multiplier $F$ depends also on a parameter $c_{\rm novel}$ that
quantifies the importance of the test. In other words, $c_{\rm novel}$
is a measure of the impact of the experiment or of the observation, that
is, how well the new observation explores novel ``dimensions'' of the
parameter and variable spaces of both the process and the model that can
reveal potential flaws. A fundamental challenge is that the
determination of $c_{\rm novel}$ requires, in some sense, a pre-existing
understanding of the physical processes so that the value of a new
experiment can be fully appreciated. In concrete situations, one has
only a limited understanding of the physical processes and the value of
a new observation is only assessed after a long learning phase, after
comparison with other observations and experiments, as well as after
comparison with the model making $c_{\rm novel}$ possibly self-referencing.
Thus, we consider $c_{\rm novel}$
is basically a judgment-based weighting of experimental referents, in which
judgment (for example, by a subject matter expert) is dominant
in its determination. The fundamental problem is to quantify the relevance of a 
new experimental referent for validation to a given decision-making problem,
given that the experimental domain of the test does
not overlap with the application domain of the decision. 
Assignment of $c_{\rm novel}$ requires the judgment of subject matter
experts, whose opinions will likely vary. This variability must be
acknowledge (if not accounted for however naively) in assigning $c_{\rm novel}$.
Thus, providing an a
priori value for $c_{\rm novel}$, as required in expression
(\ref{mgler}), remains a difficult and key step in the validation
process. This difficulty is similar to specifying the utility function
in decision making \cite{Morgen}.

Repeating an experiment twice is a special degenerate case
since it amounts ideally to increasing the statistical size of the sample.
In such a situation, one should aggregate the two
experiments 1 and 2 (yielding the relative likelihoods $p_1/q$ and $p_2/q$ respectively) 
graded with the same $c_{\rm novel}$ 
into an effective single test with the same $c_{\rm novel}$
and likelihood $(p_1/q)(p_2/q)$. This is the ideal situation, as there
are cases where repeating an experiment may wildly increase the presence of epistemic
uncertainty (or demonstrate uncontrolled variability or other kinds of
problems). When this occurs, this means that the assumption that there is no
surprise, no novelty, in repeating the experiment is incorrect. Then, 
the two experiments should be treated so as to contribute
two multipliers $F$'s, because they reveal different kinds of uncertainty
that can be generated by ensembles of experiments.

One experimental test corresponds to a entire loop $1-4$ transforming
a given $V_{\rm prior}$ to a $V_{\rm posterior}$ according to (\ref{mgler}).
This $V_{\rm posterior}$ becomes the new $V_{\rm prior}$
for the next test, which will transform it into
another $V_{\rm posterior}$ and so on,
according to the following iteration process:
\be
V_{\rm prior}^{(1)} \to
V_{\rm posterior}^{(1)} = V_{\rm prior}^{(2)} \to
V_{\rm posterior}^{(2)} = V_{\rm prior}^{(3)} \to \dots \to
V_{\rm posterior}^{(n)}~.
\label{mgmbmel}
\ee
After $n$ validation loops, we end up with a posterior
trust in the model given by
\be
V_{\rm posterior}^{(n)} / V_{\rm prior}^{(1)} =
F\left[ p^{(1)}(M | y^{(1)})_{\rm obs}, q^{(1)}; c^{(1)}_{\rm novel}\right]
~\dots~
F\left[ p^{(n)}(M | y^{(n)}_{\rm obs}), q^{(n)}; c^{(n)}_{\rm novel} \right]~,
\label{mglerdffd}
\ee
where the product is time-ordered since the sequence of values for
$c^{(j)}_{\rm novel}$
depend on preceding tests.
Validation can be said to be asymptotically satisfied when the number of
steps $n$ and the final value $V_{\rm posterior}^{(n)}$ are sufficiently high.
How high is high enough is subjective and may depend on both the application and
programmatic constraints. The concrete examples discussed below offer
some insight on this issue.
This construction makes clear that there is no absolute validation, only
a process of corroborating or disproving steps competing in a global
valuation of the model under scrutiny.
The product
(\ref{mglerdffd}) expresses the
assumption that successive observations give independent multipliers. This
assumption keeps the procedure simple because determining
the dependence between different tests with respect to validation would be
highly undetermined. We propose that it is more convenient to measure 
the dependence 
through the single parameter $c^{(j)}_{\rm novel}$ quantifying the novelty
of the $j$th test with respect to those preceding it. In full generality,
each new $F$ multiplier should be a function of all previous tests.

The loop $1-4$ together with expression (\ref{mgler}) are offered as
an attempt to
quantify the progression of the validation process, so that
eventually, when several approximately
independent tests exploring different features of the model and
of the process have
been performed, $V_{\rm posterior}$ has grown to a level at which
most experts will be satisfied
and will believe in the validity of the model. This formulation has
the advantage of
viewing the validation process as a convergence or divergence built
on a succession of steps, mimicking the construction of a theory of
reality \cite{footnoteconverg}.
Expression (\ref{mglerdffd}) embodies
the progressive build-up of trust in a model or theory.
This formulation provides a formal setting for discussing
the difficulties that underlay the so-called impossibilities 
\cite{Oreskes,auxi} in validating a given model.
Here, these difficulties are not only partitioned but quantified:
\begin{itemize}
\item
in the definition of ``new'' non-redundant experiments (parameter
$c_{\rm novel}$),
\item
in choosing the metrics and the corresponding statistical tests
quantifying the
comparison between the model and the measurements of this experiment
(leading to the likelihood ratio $p/q$), and
\item
in iterating the procedure so that the product of the gain/loss
factors $V_{\rm posterior}/V_{\rm prior}$
obtained after each test eventually leads to a clear-cut conclusion
after several tests.
\end{itemize}
This formulation makes clear why and how one is never fully convinced
that validation has
been obtained: it is a matter of degree, of confidence level, of
decision making, as in
statistical testing. But, this formulation helps in quantifying what
new confidence (or distrust) is gained in a given model.
It emphasizes that validation is an ongoing process,
similar to the never-ending construction of a theory of reality.

The general formulation proposed here in terms of iterated validation
loops is intimately linked
with decision theory based on limited knowledge: the decision to
``go ahead'' and use the model
is fundamentally a decision problem based on the accumulated
confidence embodied
in $V_{\rm posterior}$. The ``go/no-go'' decision must take into account
conflicting requirements and compromise between different
objectives. Decision theory, created by the statistician Abraham
Wald in the late forties,
is based ultimately on game theory \cite{Morgen,Kotz}.
Wald \cite{Wald} used
the term {\it loss function}, which is the standard terminology used in
mathematical statistics. In mathematical economics, the opposite of
the loss (or cost) function gives the concept of the {\it utility function},
which quantifies (in a specific functional form) what is considered
important and robust in the fit of the model to the data.
We use $V_{\rm posterior}$ in an even more general sense than ``utility,''
as a decision and information-based valuation that supports risk-informed decision-making
based on ``satisficing'' \cite{satisficing} (see the concrete examples
discussed below).

While expression (\ref{mgler}) is reminiscent of a Bayesian analysis, it
does not deal with probabilities. In the Bayesian methodology
of validation \cite{Maha1,Maha2}, only comparison between models can
be performed due to the need to remove the unknown probability of the data
in Bayes's formula. In contrast, our approach provides a value for
each single model independently of the others. In addition, it emphasizes 
the importance of quantifying the novelty
of each test and takes a more general view on how to use the
information provided from the goodness-of-fit.
The valuation (\ref{mgler}) of a model uses probabilities as partial inputs,
not as the qualifying criteria for model validation. This 
does not mean however that there are not uncertainties in these quantities 
or in the terms $F$, $q$ or $c_{\rm novel}$ and that
aleatory and systemic uncertainties \cite{Aleatory} are ignored,
as discussed below.

\vskip 0.5cm
\noindent
{\bf Properties of the Multiplier of the Validation Step}

\noindent
The multiplier $F\left[ p(M | y_{\rm obs}), q; c_{\rm novel} \right]$
should have the following properties:
\begin{enumerate}
\item
If the statistical test(s) performed on the given observations
is (are) passed at the reference level $q$, then the posterior
potential value is larger
than the prior potential value: $F>1$ (resp. $F \leq 1$) for $p>q$
(resp. $p \leq q$), which can be written succinctly as $\ln F / \ln (p/q) > 0$.
\item
The larger the statistical significance of the passed test,
the larger the posterior value. Hence
\be
{\partial F \over \partial p} > 0~,
\ee
for a given $q$.
There could be a saturation of the growth of $F$ for large $p/q$,
which can be either that $F < \infty$ as $p/q \to \infty$ or of the form
of a concavity requirement $\partial^2 F / \partial p^2 <0$ for
large $p/q$: obtaining a quality of fit beyond a certain level should not be attempted.
\item
The larger the statistical level at which the test(s)
performed on the given observations
is (are) passed, the larger the impact of a ``novel'' experiment on
the multiplier
enhancing the prior into the posterior potential value of the model:
$\partial F / \partial c_{\rm novel} > 0$ (resp. $\leq 0$), for
$p>q$ (resp. $p \leq q$).
\end{enumerate}

The simplest form obeying these properties (not including the saturation
of the growth of $F$) is
\be
F\left[ p(M | y_{\rm obs}), q; c_{\rm novel} \right] =
\left( {p \over q} \right)^{c_{\rm novel}}~.
\label{mfmsl}
\ee
This form provides an intuitive interpretation of the meaning of the
experiment impact parameter $c_{\rm novel}$.
A bland evaluation of the novelty of a test would be $c_{\rm novel} = 1$,
thus $F = p/q$ and
the chain (\ref{mglerdffd}) reduces to a product of normalized likelihoods, as
in standard statistical tests. A value $c_{\rm novel} > 1$ (resp. $< 1$)
for a given experiment describes
a nonlinear rapid (resp. slow) updating of our trust $V$ as a function of the
grade $p/q$ of the model with respect to the observations. In particular,
a large value of $c_{\rm novel}$ corresponds to the case of ``critical'' tests.
A famous example is the Michelson-Morley experiment for the Theory of
Special Relativity.
For the Theory of General Relativity, it was the observation during the 1919
solar eclipse of the bending of light rays from distant stars by the
Sun's mass and the anomalous precession of the perihelion of
Mercury's orbit. 

Note that the parameterization (\ref{mfmsl}) should account for the
decreased novelty noted above occurring when the 
same experiment is repeated two or more times. The value of 
$c_{\rm novel}$ should be reduced for each repetition of the same test;
moreover, the value of $c_{\rm novel}$ should approach unity as the number
of repetitions increases.

\begin{figure}
\includegraphics[width=14cm]{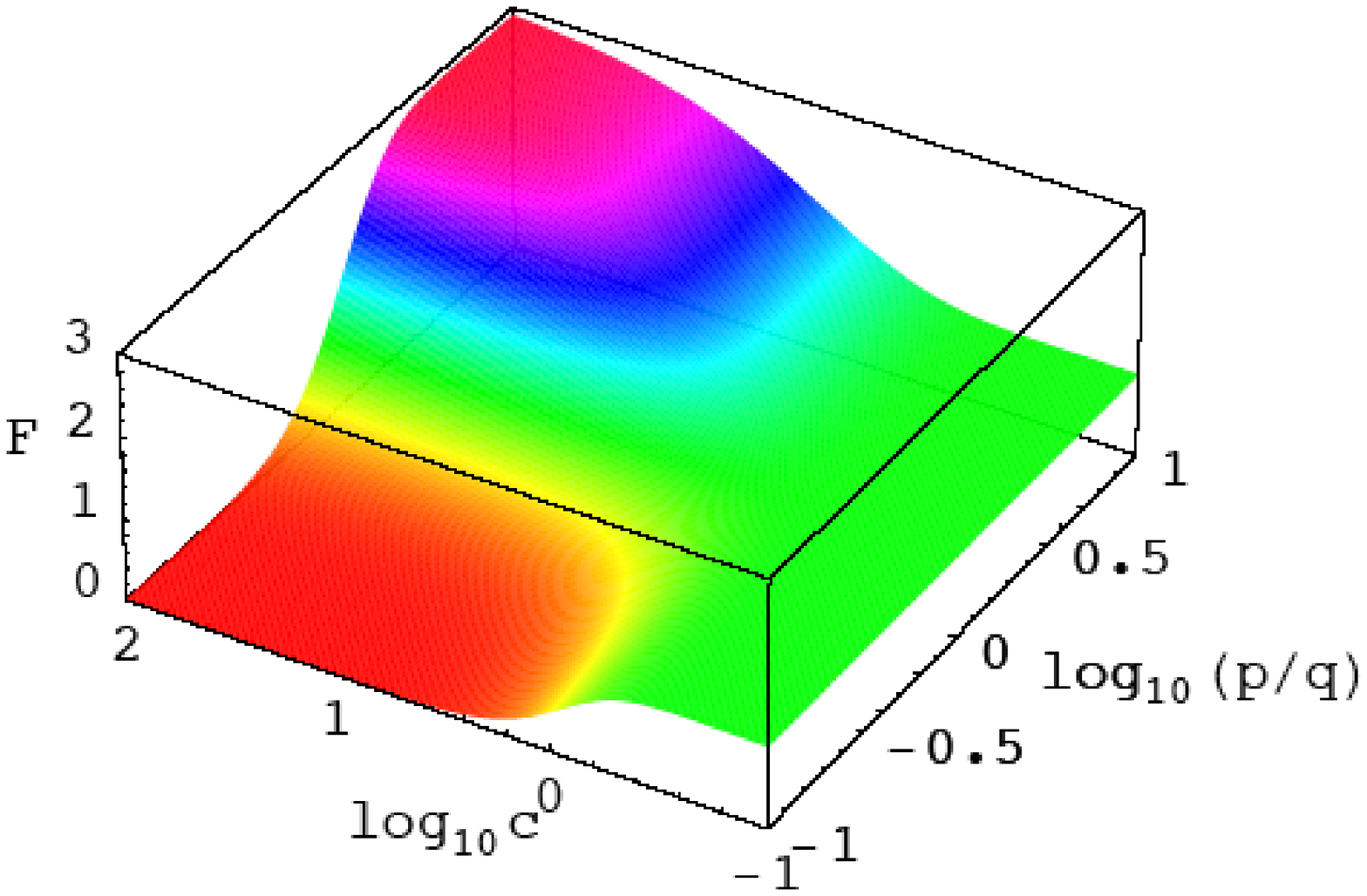}
\caption{ The multiplier defined by
(\ref{mgler222}) is plotted as a function of $p/q$ and $c_{\rm novel}$.
}
\label{Ftanh}
\end{figure}

The alternative multiplier,
\be
F\left[ p(M | y_{\rm obs}), q; c_{\rm novel} \right] =
\left[{\tanh \left( {p \over q} + {1 \over c_{\rm novel}} \right) \over
\tanh \left( 1+ {1 \over c_{\rm novel}} \right)}\right]^4~,
\label{mgler222}
\ee
is plotted in Fig.~\ref{Ftanh} as a function of $p/q$ and $c_{\rm novel}$.
It emphasizes that $F$ saturates as a function of
$p/q$ and $c_{\rm novel}$ as either one or both of them grow large.
A completely new
experiment corresponds to $c_{\rm novel} \to \infty$ so that
$1/c_{\rm novel}=0$ and
thus $F$ tends to $[\tanh(p/q)/\tanh(1)]^4$, i.e.,
$V_{\rm posterior}/V_{\rm prior}$ is
only determined by the quality of the ``fit'' of the data by the model
quantified by $p/q$.  A finite $c_{\rm novel}$ implies that
one already takes a restrained view on the usefulness of the
experiment since one limits the amplitude of the
${\rm gain}=V_{\rm posterior}/V_{\rm prior}$, whatever the quality of
the fit of the data by
the model. The exponent $4$
in (\ref{mgler222}) has been
chosen so that the maximum confidence gain $F$ is equal to
$1/(\tanh(1))^4 \approx 3$ in
the best possible situation of a completely new experiment
($c_{\rm novel} = \infty$) and perfect fit ($p/q \to \infty$).
In contrast, the multiplier $F$ can be arbitrarily small as $p/q \to 0$
even if the novelty of the test is high ($c_{\rm novel} \to \infty$).
For a finite novelty $c_{\rm novel}$,
a test that fails the model miserably ($p/q \approx 0$) does not necessarily
reject the model completely: unlike with the expression in
(\ref{mfmsl}), $F$ remains greater than zero.  Indeed, if the novelty $c_{\rm
novel}$ is small, the worst-case multiplier (attained for $p/q = 0$)
is $\left[
\tanh \left( 1/c_{\rm novel} \right) /
\tanh \left( 1+(1/c_{\rm novel}) \right)
\right]^4 \approx 1 - 6.9\,{\rm e}^{-2/c_{\rm novel}}$,
which is only slightly less than unity if $c_{\rm novel} \ll 1$.
In short, this formulation does not heavily weight unimportant tests.

In the framework of decision theory, expression (\ref{mgler}) with one of the
specific expressions in (\ref{mfmsl}) or (\ref{mgler222})
provides a parametric form for the utility or decision ``function''
of the decision maker. It is clear
that many other forms of the utility function can be used, however, with the
constraint of keeping the salient features of expression (\ref{mgler})
with (\ref{mfmsl}) or (\ref{mgler222}),
in terms of the impact of a new test given past tests, and the quality of the
comparison between the model predictions and the data.

Finally, we remark that the proposed form for the multiplier (\ref{mgler222})
contains an important asymmetry between gains and losses: the
failure to a single test with strong novelty and significance (as, 
e.g., for the
localized seismicity on faults in the case of the OFC model and
for the leverage effect in the case of the MRW model discussed below) cannot be compensated
by the success of all the other tests combined. In other words,
a single test is enough to reject a model.
This embodies the common lore that reputation gain is a slow process requiring
constancy and tenacity,
while its loss can occur suddenly with one single failure
and is difficult to re-establish. We believe that the
same applies to the build-up of trust in and, thus, validation of a model.

\vskip 0.5cm
\noindent
{\bf Practical Guidelines for Determining $p/q$ and $c_{\rm novel}$}

\noindent
These two crucial elements of a validation step are conditioned by four
basic problems, over which one can exert at least partial control.
In particular, they address the two sources of uncertainty: 
systemic (lack of knowledge, important missing mechanisms) and aleatory \cite{Aleatory}
(due to variability inherent in the phenomenon under consideration).
In a nutshell, as becomes clearer below, the comparison between $p$ and $q$
is more concerned with the aleatory uncertainty
while $c_{\rm novel}$ deals in part with the systemic uncertainty.
In the following, as in the two examples (\ref{mfmsl}) and
(\ref{mgler222}), we 
consider that $p$ and $q$ enter only in the form of their ratio $p/q$.
This should not be generally the case but, given the many uncertainties, 
this restriction seems to simplifly the analysis by removing a degree
of freedom.

\begin{enumerate}
\item
{\it How to model?}
This addresses model construction and involves the structure of the
elementary contributions, their hierarchical organization, and requires dealing with
uncertainties and fuzziness. This concerns the
epistemic uncertainty.
\item
{\it What to measure?}
This relates to the nature of $c_{\rm novel}$: ideally,
following Palmer et al. \cite{Palmer}, one should
target adaptively the observations to ``sensitive'' parts of the system.
Targeting observations could be directed by the
desire to access the most ``relevant'' information
as well as to get information that is the most reliable, i.e., which is
contaminated by the smallest errors. This is also the stance
of Oberkampf and Trucano \cite{Ober}: ``A validation experiment is
conducted for the primary purpose
of determining the validity, or predictive accuracy, of a computational
modeling and simulation capability. In other words, a validation
experiment is designed, executed, and analyzed for the purpose of
quantitatively determining the ability of a mathematical model and its
embodiment in a computer code to simulate a well-characterized physical
process.'' In practice, $c_{\rm novel}$ is chosen to represent the
best guess-estimate
of the importance of the new observation and the degree of
``surprise'' it brings to the validation step \cite{Kevin}.
The epistemic uncertainty alluded to above is partially addressed
in the choice of the empirical data and its rating $c_{\rm novel}$
(see the examples of application discussed below).
\item
{\it How to measure?}
For given measurements or experiments, the problem is to
find the ``optimal'' metric or cost function
(involved in the quality-of-fit measure $p$)
for the intended use of the model. The notion of
optimality needs to be defined. It could capture a compromise between
fitting best the ``important'' features of the data (what is
``important'' may be
decided on the basis of previous studies and understanding or other
processes, or
programmatic concerns), and minimizing the extraction of spurious
information from noise. This requires one
to have a precise idea of the statistical properties of the noise. If such
knowledge is not available, the cost function should be chosen accordingly.
The choice of the ``cost function'' involves the choice of how to
look at the data.
For instance, one may want to expand the measurements at multiple scales using
wavelet decompositions and compare the prediction and observations
scale by scale, or
in terms of multifractal spectra of the physical fields estimated from these
wavelet decompositions \cite{Muzyetal94} or from other methods.
The general idea here is that, given complex observation fields,
it is appropriate to unfold the data on a variety of
``metrics,'' which can then be used
in the comparison between observations and model predictions: the question is
then how well is the model able to reproduce the salient
multiscale and multifractal
properties derived from the observations? The physics of turbulent
fields and of complex
systems have offered many such new tools with which to
unfold complex fields according to different statistics.
Each of these statistics offers a metric to compare
observations with model predictions and is associated with a cost function
focusing on a particular feature of the process. Since these metrics
are derived
from the understanding that turbulent fields can be analyzed using
these metrics
that reveal strong constraints in their organization, these metrics can
justifiably be called ``physics-based.''
In practice, $p$, and eventually $p/q$, has to be inferred
as an estimate of the degree of matching between the model output and the
observation. This can be done following the concept of fuzzy logic in which one
replaces the yes/no pass test by a more gradual quantification of
matching \cite{fuzzy}. We thus concur with Ref.~\cite{Oberkampf2005},
while our general methodology goes beyond. Note that this 
discussion relates to the aleatory uncertainty \cite{Aleatory}.

\item
{\it How to interpret the results?}
This question relates to defining the test and the reference probability
level $q$ that any other model (than the one under scrutiny) can explain
the data. The interpretation of the results should aim at detecting the
``dimensions'' that are missing, mispecified or erroneous in the model
(systemic uncertainty).
What tests can be used to betray the existence of hidden degrees of
freedom and/or dimensions? This is the hardest problem. It can sometimes
find an elegant solution when a given model is embedded in a more
general one. Then, the limitation of the ``smaller'' model becomes clear
from the vantage of the more general model.
\end{enumerate}

\noindent
We now illustrate our algorithmic approach to model validation using 
the historical development of quantum mechanics and four examples 
based on the authors' research activities.
In these examples, we will use the form (\ref{mgler222}) and consider 
three finite values:
$c_{\rm novel}=1$ (marginally useful new test),
$c_{\rm novel}=10$ (substantially new test), and
$c_{\rm novel}=100$ (important new test).
When a likelihood test is not available,
we propose to use three possible marks:
$p/q=0.1$ (poor fit),
$p/q=1$ (marginally good fit), and
$p/q=10$ (good fit).
Extreme values ($c_{\rm novel}$ or $p/q$ are 0 or $\infty$) have
already been discussed.  Due to limited experience with
this approach, we propose these ad hoc values in the following
examples of its application.

\vskip 0.5cm
\noindent
{\bf Quantum Mechanics}

\noindent
Quantum mechanics (QM)
offer a vivid incarnation of how a model can turn progressively into a
theory held ``true'' by almost all physicists. Since its birth, QM has
been tested again and again because it presents a view of ``reality''
that is shockingly different from the classical view experienced at the
macroscopic scale. QM prescriptions and predictions often go against classical
intuition. Nevertheless, we can state that, by a long and thorough
process of verified predictions of QM in experiments,
fueled by the imaginative set-up of paradoxes, QM has been validated as
a correct description of nature.
It is fair to say that the overwhelming majority of physicists have
developed a strong trust in the validity of QM. That is, if someone comes up
with a new test based on a new paradox, for instance, most physicists would
bet that QM will come up with the right answer with a very high
probability. It is thus by the on-going testing and the compatibility of
the prediction of QM with the observations that QM has been validated.
As a consequence, one can use it with strong confidence to make
predictions in novel directions. This is ideally the situation one would
like to attain for the problem of validation of models 
discussed below. We now give a very partial list of selected
tests that established the trust of physicists in Quantum
Mechanics.

\begin{enumerate}

\item
Pauli's exclusion principle states that no two identical fermions 
(particles with non-integer values of spin) may
occupy the same quantum state simultaneously \cite{Pauliprinciple}. 
It is one of the most
important principles in physics, primarily because the three types of
particle from which ordinary matter is made, electrons, protons, and
neutrons, are all subject to it. With $c_{\rm novel} = 100$
and perfect agreement in numerous experiments ($p/q = \infty$), 
this leads to
$F^{(1)}=2.9$.

\item
The EPR paradox \cite{EPR} was a thought experiment designed to prove that
quantum mechanics was hopelessly flawed: according to QM,
a measurement performed on one part of a quantum system
can have an instantaneous effect on the result of a measurement
performed on another part, regardless of the distance separating the two
parts. Bell's theorem \cite{Bell} showed that quantum mechanics 
predicted stronger
statistical correlations between entangled particles than the so-called
local realistic theory with hidden variables. The importance of this prediction
requires $c_{\rm novel} = 100$ at the very minimum. The QM prediction 
turned out to be correct,
winning over the hidden-variables theories \cite{Aspect,Rarity} ($p/q 
= \infty$),
leading again to $F^{(2)}=2.9$.

\item
The Aharonov-Bohm effect predicts that a magnetic field can
influence an electron that, strictly speaking, is located completely beyond the
field's range, again an impossibility according to non-quantum theories
($c_{\rm novel} = 100$).
The Aharonov-Bohm oscillations were observed
in ordinary (i.e., nonsuperconducting) metallic rings, showing that 
electrons can maintain
quantum mechanical phase coherence in ordinary materials \cite{Webb}.
This yields $p/q = \infty$ and thus $F^{(3)}=2.9$ yet again.

\item
The Josephson effect provides a macroscopic incarnation of quantum
effects in which two superconductors are predicted to preserve their
long-range order across an insulating barrier, for instance, leading to rapid
alternating currents when a steady voltage is applied accross the 
superconductors.
The novelty of this effect again warrants $c_{\rm novel} = 100$ and the
numerous verifications and applications (for instance in SQUIDs:
{S}uperconducting {QU}antum {I}nterference {D}evices) argues for $p/q = \infty$
and thus $F^{(4)}=2.9$, as usual.

\item
The prediction of possible collapse of a gas of atoms at low temperature
into a single quantum state is known as
Bose-Einstein (BE) condensation, again so much against classical intuition
($c_{\rm novel} = 100$). Atoms are indeed bosons (particles with 
integer values of spin), which are {\it not} subjected to the Pauli 
exclusion principle evoked in the above test \#1 of QM.  The first such 
BE condensate was produced
using a gas of rubidium atoms cooled to $1.7 \cdot 10^{-7}$ K \cite{weiman}
($p/q = \infty$), leading once more to $F^{(4)}=2.9$.

\item
There have been several attempts to develop a paradox-free
nonlinear QM theory, in the hope of eliminating Schr\"odinger's cat
paradox, among other embarrassments. The nonlinear QM
predictions diverge from those of orthodox quantum physics, albeit
subtly. For instance, if a neutron impinges on two slits, an
interference pattern appears, which should, however, disappear if the 
measurement
is made far enough away ($c_{\rm novel} = 100$). Experiment tests
of the neutron prediction rejected the nonlinear version in favor
of the standard QM \cite{Gahler} ($p/q = \infty$), leading to $F^{(6)}=2.9$.

\item
In addition, measurements at the National Bureau of Standards in 
Boulder, CO, on frequency standards
have been shown to set limits of
order $10^{-21}$ on the fraction of the energy of the rf
transition in $^9$Be ions that could
be due to nonlinear corrections to quantum mechanics \cite{Weinberg}.
We assign $c_{\rm novel} = 10$, with $p/q = 10$), to this result, 
leading to $F^{(7)}=2.4$.  Although less than $F^{(1-6)}$ this is 
still meant to be an impressive score.

\end{enumerate}
\noindent
Combining the multipliers according to (\ref{mglerdffd})
leads to $V_{\rm posterior}^{(8)} / V_{\rm prior}^{(1)} \simeq 1400$, which
is of course only a lower limit given the many other validation tests
not mentioned here. 

Tests of QM are ongoing \cite{Leggett}.
But given the presumably huge amount of trust physicists have in QM
which we tried to quantify,
why do physicists still feel the need to put QM to the ``validation test?'' 
This raises the question whether we can ever establish a sense
of sufficiency for validation. Our position is that this reflects
the quixotic quest for the absolute truth, and also the taste for surprises.
Perhaps, by continuing to test QM, humans will uncover a new insight or an
anomaly which may help progress in the understanding of reality.

\vskip 0.5cm
\noindent
{\bf Four Further Examples Drawn from the Authors' Research Activities}
\vskip 0.3cm
\noindent
{\it The Olami-Feder-Christensen (OFC) sand-pile model of earthquakes.}
This is perhaps the simplest sand-pile model of self-organized criticality,
which exhibits a phenomenology resembling real seismicity \cite{OFC}.
Figure 2 shows a ``stress'' map generated by the OFC model
immediately after a large avalanche (main shock)
at two magnifications, to illustrate the rich organization of almost synchronized
regions \cite{Drossel}. To validate the OFC model,
we examine the properties and prediction of the model that can be compared
with real seismicity, together with our assessment of their $c_{\rm novel}$
and quality-of-fit. We are careful to state these properties in an
ordered way, as specified in the above sequences 
(\ref{mgmbmel})--(\ref{mglerdffd}).
\begin{enumerate}
\item
The statistical physics community recognized the discovery of the OFC model as an important
step in the development of a theory of earthquakes: without a conservation law
(which was thought before to be an essential condition), it 
nevertheless exhibits
a power law distribution of avalanche sizes resembling the
Gutenberg-Richter (GR) law \cite{OFC}. On the other hand,
many other models with different mechanisms can explain observed 
power law distributions
\cite{bookcritical}. We thus attribute only $c_{\rm novel} = 10$ to 
this evidence.
Because the power law distribution
obtained by the model is of excellent quality
for a certain parameter value ($\alpha \approx 0.2$),
we formally take $p/q = \infty$ (perfect fit).
Expression (\ref{mgler222}) then gives $F^{(1)}=2.4$.

\item Prediction of the OFC model concerning
foreshocks and aftershocks, and their exponents for the inverse
and direct Omori laws. These predictions are twofold \cite{HHS}:
(i) the finding of foreshocks and aftershocks with similar
qualitative properties,
and (ii) their inverse and direct Omori rates. The first aspect,
deserves a large $c_{\rm novel}=100$
as the observation of foreshocks and aftershocks came as a rather big surprise
in such sand-pile models \cite{StephanPRL}. The clustering in time and space
of the foreshocks and aftershocks are qualitatively similar to real
seismicity \cite{HHS},
which warrants $p/q=10$, and thus $F^{(2a)}=2.9$. The second aspect 
is secondary
compared with the first one ($c_{\rm novel}=1$).
Since the exponents are only qualitatively reproduced (but
with no formal likelihood test available), we take $p/q=0.1$.
This leads to $F^{(2b)}=0.47$.
\item
Scaling of the number of aftershocks with the main shock size
(productivity law) \cite{HHS}:
$c_{\rm novel}=10$ as this observation is rather new but not
completely independent
of the Omori law. The fit is good so we grant a grade $p/q=10$ leading to
$F^{(3)}=2.4$.
\item
Power law increase of the number of foreshocks with the
mainshock size \cite{HHS}:
this is not observed in real seismicity, probably because this
property is absent
or perhaps due to a lack of quality data. This test is therefore not
very selective ($c_{\rm novel}=1$) and the large uncertainties suggest
a grade $p/q=1$ (to reflect the different viewpoints on the
absence of effect in real data) leading to $F^{(4)}=1$ (neutral test).
\item
Most aftershocks are found to nucleate at ``asperities'' located on the
mainshock rupture plane or on the boundary of the avalanche, in agreement with
observations \cite{HHS}: $c_{\rm novel}=10$ and $p/q=10$ leading to
$F^{(5)}=2.4$.
\item
Earthquakes cluster on spatially localized geometrical structures,
known as faults. This property is arguably central to the physics of seismicity 
($c_{\rm novel}=100$), but absolutely not reproduced by the 
OFC model
($p/q=0.1$). This leads to $F^{(6)}= 4 \cdot 10^{-4}$.
\end{enumerate}
Combining the multipliers according to (\ref{mglerdffd}) up to test \#5
leads to $V_{\rm posterior}^{(6)} / V_{\rm prior}^{(1)} =  18.8$,
suggesting that the OFC model is validated as a useful model of the
statistical properties of seismic catalogs, at least with respect to the
properties which have been examined in these first five tests. Adding the
crucial last test strongly fails the model since $V_{\rm posterior}^{(7)} /
V_{\rm prior}^{(1)} = 0.0075$. The model can not be used as a realistic
predictor of seismicity. It can nevertheless be useful to illustrate
certain statistical properties and to help formulate new questions
and hypotheses.

\vskip 0.3cm
\noindent
{\it The multifractal random walk (MRW) as a model of financial returns}.
We now consider the MRW model introduced as
a random walk with stochastic ``volatility'' endowed with exact
multifractal properties \cite{M_etal}, which has been proposed as a model
of financial time series.
Among the documented facts about financial time series, we have
the absence of correlation between lagged returns,
the long-range correlation of lagged volatilities, and the observed
multifractality.
These can not be taken as validation tests of the model since they are the
observations that motivated the introduction of the MRW.
These observations thus constitute
references or benchmarks against which new tests must be compared.
The new properties and prediction of the MRW model that can be compared
with real financial return time series are the following.
\begin{enumerate}
\item
The probability density distributions (PDF) of returns at different
time scales (see Figure 3):
the MRW exhibits the remarkable property of accounting quantitatively
for the transition from fatter-than-exponential PDFs at small time scales
to approximately Gaussian PDFs at large time scales. But, because the MRW
is intrinsically a model developed as the continuous limit of a cascade
across scales, this is perhaps not very surprising. We thus rate the
novelty of this observation with $c_{\rm novel}=10$. In absence of
formal likelihood tests on the PDFs, we take $p/q=10$ to reflect the
apparent excellent fits of the data at multiple scales, leading to
$F^{(1)}=2.4$.
\item
Different response functions of the price volatility
to large external shocks compared with endogeneous shocks,
which are well-confirmed quantitatively by observations
on a hierarchy of volatility shocks \cite{SorMarMuzy}. This
prediction has been verified
to hold with remarkable accuracy without any adjustable parameters
(i.e., the parameters were adjusted previously and fixed before the new test).
We thus rate the novelty of this test with a high $c_{\rm novel}=100$ and the
agreement is quantified by $p/q=10$, leading to
$F^{(2)}=2.9$.
\item
The sharp-peak/flat-trough
pattern of price peaks \cite{RoehnerSorSP} as well as
accelerated speculative bubbles preceding crashes \cite{JSL} is
not captured by the MRW. In view of the debated importance of such
patterns, we rate these observations with $c_{\rm novel}=1$ and
$p/q=0.1$, leading to  $F^{(3)}=0.47$.
\item
The leverage effect and volatility dependence on past
volatility and returns (see
\cite{bouetal} and references therein).
These features are not captured by the MRW at all.  We rate $c_{\rm
novel}=10$ and the
lack of agreement is quantified by $p/q=0.1$, leading to
$F^{(4)}=0.0037$.
\end{enumerate}
Combining the multipliers according to (\ref{mglerdffd}) leads to
$V_{\rm posterior}^{(5)} / V_{\rm prior}^{(1)} = 0.012$, rejecting the model.
But if we stop the
validation steps at $V_{\rm posterior}^{(3)} / V_{\rm prior}^{(1)} = 7$, we
obtain a clear validation signal. The two additional tests fail the MRW
because the observed effects involve mechanisms that are absent in
it. Here, we should
conclude that the MRW is a useful model that is validated with respect to
certain properties on the memory of volatility but is not validated for a
fully faithful description of the stock market returns. These mechanisms can be
actually incorporated into extensions of the MRW, corresponding
to the addition of new dimensions lacking in the MRW.
If we had used the long-range correlation
of lagged volatilities and the observed multifractality (each with parameters
$c_{\rm novel}=10$ and $p/q=10$) as tests \#-1 and \#0, $F$ would 
have gained a factor $2.4^2 = 5.9$,
changing $V_{\rm posterior}^{(5)} / V_{\rm prior}^{(1)} = 0.012$ into
$V_{\rm posterior}^{(5)} / V_{\rm prior}^{(-1)} = 0.07$, still far 
from sufficient to validate the model.

\vskip 0.3cm
\noindent
{\it An anomalous diffusion model for solar photons in cloudy atmospheres.}
To properly model climate dynamics, it is important to reduce the
significant uncertainty associated with clouds. In particular,
estimation of the radiation budget in the presence of clouds needs to
be improved since current operational models for the most part ignore all
variability below the scale of the climate model's grid (a few
100~km).  So a considerable effort has been expended to derive more
realistic mean-field radiative transfer models \cite{Barker05},
mostly by considering only the one-point variability of clouds (that
is, irrespective of their actual structure).  However, it has been
widely recognized that the Earth's cloudiness is fractal over a wide 
range of scales \cite{Lovejoy}. This is
the motivation for modeling the paths of solar photons at
non-absorbing wavelengths in the cloudy atmosphere as convoluted
L\'evy walks \cite{bookcritical}, which are characterized by frequent
small steps (inside clouds) and occasional large jumps (typically between
clouds) as represented schematically in Fig.~\ref{anomalous_diffusion}. 
These paths start downward at the top of the highest clouds
and end in escape to space or in absorption at the surface. In sharp
contrast with most other mean-field models for solar radiative
transfer, this diffusion model with anomalous scaling can be
subjected to a battery of observational tests.
\begin{enumerate}
\item
The original goal of this phenomenological model, which accounts for
the clustering of cloud water droplets into broken and/or
multi-layered cloudiness, was to predict the increase in
steady-state flux transmitted to the surface compared to what would
filter through that same amount of water in a single unbroken cloud
layer \cite{Davis97}.  This property is common to all mean-field
photon transport models that do anything at all about unresolved
variability \cite{Barker05}. Thus, we assign only $c_{\rm novel}=10$
to this test and, given that all models in this class are successful,
we have to take $p/q=1$, hence $F^{(1)}=1$.  The outcome of this
first test is neutral.
\item
The first real test for this model occurred when it became possible
to accurately estimate the mean total path of solar photons that
reach the surface.  This breakthrough was enabled by access to
spectroscopy at medium (high) resolution  of oxygen bands (lines)
\cite{Pfiel99,Min01}.  Along with simultaneous estimation of cloud
optical depth (basically, column-integrated water [kg/m$^2$] times
the average scattering cross-section per kg), the observed trends were
explained only by the new model in spite of the relatively large
instrumental error bars.  So we assign $c_{\rm novel}=100$ to this
highly discriminating test and $p/q=10$ (even though other models
were generally not in a position to compete), hence $F^{(2)}=2.9$.
\item
Another test was proposed using time-dependent photon transport with
a source near the surface (cloud-to-ground lightning) and a detector
in space (the DOE FORT\'{E} satellite) \cite{Davis00}.  The quantity
of interest is the observed delay of the light pulse (due to
multiple scattering in the cloud system) with respect to the
radio-frequency pulse (which travels in a straight line).  There was
no simultaneous estimate of cloud optical depth, so assumptions had
to be made (informed by the fact that storm clouds are at once thick
and dense).  Because of this lack of an independent measurement, we
assign only $c_{\rm novel}=10$ to the observation and $p/q=1$ to the
model performance since this is only about the finite horizontal
extent of the cloud (one could exclude only uniform
``plane-parallel'' clouds).  So, again we obtain $F^{(3)}=1$ for an
interesting but presently neutral test that needs to be refined.
\item
Min et al. \cite{Min04} developed an oxygen-line spectrometer with
sufficient resolution to estimate not just the {\it mean} path but
also its {\it root-mean-square} (RMS) value.  They found the
prediction by Davis and Marshak \cite{Davis02} for normal diffusion
to be an extreme (envelop) case for the empirical scatter plot of
mean vs. RMS path, and this is indicative that the anomalous
diffusion model will cover the bulk of the data.  Because of some
overlap with a previous item, we assign $c_{\rm novel}=10$ and
$p/q=10$ for the model performance (since the anomalous diffusion model had
not yet made a prediction for the RMS path, but other models have
yet to make one for the mean path). We therefore obtain $F^{(4)}=2.4$.
\item
Using similar data but a different normalization than Min et al.'s, 
more amenable to model testing, Scholl et al.
\cite{Scholl05} observed that the RMS-to-mean ratio for solar photon
path is essentially constant whether the diffusion is normal or
anomalous.  This is a remarkable empirical finding to which we assign
$c_{\rm novel}=100$.  The new mean- and RMS-path data was explained
by Scholl et al. by creating an ad hoc hybrid between the normal
diffusion theory (which made a prediction for the RMS path) and its
anomalous counterpart (which did not).  This significant modification of
the basic model means that we are in principle back to validation step 1 with
the new model.  However, this exercise uncovered something quite telling
about the original anomalous diffusion model, namely, that its simple
asymptotic (large optical depth) form used in all the above tests is 
not generally valid:
for typical cloud covers, the pre-asymptotic terms computed
explicitly for the normal diffusion case prove to be important
irrespective of whether the diffusion is normal or not.
Consequently, in its original form (a simple scaling law for the
mean path with respect to cloud thickness and optical depth), the
anomalous diffusion model fails to reproduce the new data even for
the mean path. (This means that previous fits yielded ``effective'' 
anomaly parameters
and were misleading if taken literally.)
So we assign $p/q=0.1$ at best for the original model, hence
$F^{(5)}=0.0004$.
\end{enumerate}
Thus, $V_{\rm posterior}^{(6)} / V_{\rm prior}^{(1)} = 0.003$, a
fatal blow for the anomalous diffusion in its simple asymptotic form,
even though $V_{\rm posterior}^{(5)} / V_{\rm prior}^{(1)} = 7.0$
which would have been interpreted as close to a convincing
validation.  Of course, this is not the end of the story.  The
original model has already spawned Scholl et al.'s empirical hybrid
and there is a formalism based on integral (in fact,
pseudo-differential) operators that extends the anomalous {\it
diffusion} model to pre-asymptotic regimes \cite{Buldyrev01}.  More
recently, a model for anomalous {\it transport} (i.e., where angular
details matter) has been proposed that fits all of the new oxygen
spectroscopy results \cite{Davis05}.

In summary, the first and simplest incarnation of the anomalous diffusion
model for solar photon transport ran its course and demonstrated the
power of oxygen-line spectroscopy as a test for the perfomance of solar
radiative transfer models required
in climate modeling for large-scale average properties.
Eventually, new and interesting tests will become feasible when we
obtain dedicated oxygen-line spectroscopy from space (with NASA's
Orbiting Carbon Observatory mission planned for launch in 2007).
Indeed, we already know that the
asymptotic scaling for reflected photon paths \cite{Davis99} is different
from their transmitted counterparts \cite{Davis02} in both mean and
RMS. 

\vskip 0.3cm
\noindent
{\it A computational fluid dynamics (CFD) model for shock-induced
mixing and shock-tube tests.}
So far, our examples of models for complex phenomena have hailed from
quantum and statistical physics.  In the latter case, they are 
stochastic models composed of: (1)
simple code (hence rather trivial verification procedures) to
generate realizations, and (2) analytical expressions for the
ensemble-average properties (that are used in the above validation
exercises).
We now turn to gas dynamics codes which have a broad range of
applications, from astrophysical and geophysical flow simulation to
the design and performance analysis of engineering systems.
Specifically, we discuss the validation of the
``Cuervo'' code developed at  Los Alamos National
Laboratory \cite{CuervoRef}. This software, which generates solutions of 
the compressive Euler equations, have been verified against a suite
of test problems having closed-form solutions; as clearly pointed out
by Oberkampf and Trucano \cite{Ober}, however, this differs from and
also does not guarantee validation against experimental data.
A standard test case involves the Richtmyer-Meshkov (RM) instability
\cite{Richtmyer,Meshkov}, which arises when a density gradient in a
fluid is subjected to an impulsive acceleration, e.g., due to passage
of a shock wave (see Fig.~\ref{ShockTubeFig_Kamm_new}).
Evolution of the RM instability is nonlinear and hydrodynamically
complex and hence defines an excellent problem-space to assess CFD
code performance.

In the series of shock-tube experiments described in
\cite{Benjamin}, RM dynamics are realized by preparing one or more
cylinders with approximately identical axisymmetric Gaussian
concentration profiles of dense sulfur hexaflouride (SF$_6$) in air.
This (or these) vertical ``gas cylinder(s)'' is (are) subjected to a
weak shock ---Mach number $\approx$1.2--- propagating horizontally.
The ensuing dynamics are largely governed by the mismatch of the density
gradient between the gases (with the density of SF$_6$ approximately
five times that of air) and the pressure gradient through the shock
wave;  this mismatch acts as the source for baroclinic vorticity
generation.  The visualization of the density field is obtained using
a planar laser-induced fluorescence (PLIF) technique, which provides
high-resolution quantitative concentration measurements.  The
velocity field is diagnosed using particle image velocimetry (PIV),
based on correlation measurements of small-scale particles that are
lightly seeded in the initial flow field.
Careful post-processing of images from 130~$\mu$s to 1000~$\mu$s
after shock passage yields planar concentration and velocity with error bars.
\begin{enumerate}
\item
The RM flow is dominated at early times by a vortex pair (per gas cylinder).
Later, secondary instabilities rapidly transition the flow to a mixed state.
We rate $c_{\rm novel}=10$ for the observations of these two
instabilities. The Cuervo code correctly
captures these two instabilities, best observed and modeled with a
single cylinder. At this qualitative level, we rate
$p/q=10$ (good fit), which leads to $F^{(1)}=2.4$.
\item
Older data for two-cylinder experiments acquired with a fog-based
technique (rather than PLIF) showed two separated spirals
associated with the primary instability, but
the Cuervo code predicted the existence of a material
bridge.  This previously unobserved connection was experimentally 
diagnosed with the improved observational
technique.
Using $c_{\rm novel}=10$ and $p/q=10$ yields
$F^{(2)}=2.4$.
\item
The evolution of the total power as a function of time offers another
useful metric. The numerical simulation quantitatively accounts for the
exponential growth of
the power with time, within the experimental error bars. Using
$c_{\rm novel}=10$ and $p/q=10$ yields
$F^{(3)}=2.4$.
\item
The concentration power spectrum as a function of wavenumber
for different
times provides another way (in the Fourier domain) to present
the information of the hierarchy of structures already visualized
in physical space ($c_{\rm novel}=1$). The Cuervo
code correctly accounts for the low wavenumber part of the spectrum but
underestimates the high wavenumber part (beyond the deterministic-stochastic
transition wavenumber) by a factor 2 to 5. We capture this by setting
$p/q=0.1$, which yields
$F^{(4)}=0.47$.
\end{enumerate}
Combining the multipliers according to (\ref{mglerdffd}) leads to
$V_{\rm posterior}^{(5)} / V_{\rm prior}^{(1)} = 6.5$, a significant gain,
but still not sufficient to compellingly validate the Cuervo code for 
inviscid shock-induced hydrodynamic instability simulations. Intricate
experiments with three gas cylinders have been performed \cite{Kumar} 
and others are currently under way to
further stress CFD models.

\vskip 0.3cm

These examples illustrate the utility of representing the validation
process as a succession of steps, each of them characterized
by the two parameters $c_{\rm novel}$ and $p/q$.
The determination of $c_{\rm novel}$ requires expert judgment and
that of $p/q$ a careful statistical analysis, which is beyond the
scope of the present report (see Ref. \cite{Oberkampf2005} for a
detailed case study). The parameter $q$ is ideally imposed as a confidence
level, say $95\%$ or $99\%$ as in standard statistical tests. In practice,
it may depend on the experimental test and requires a case-by-case
examination.

The uncertainties of $c_{\rm novel}$ and
of $p/q$ need to be assessed. Indeed, different statistical estimations
or metrics may yield different $p/q$'s and different experts
will likely rate differently the novelty $c_{\rm novel}$ of a new 
test. As a result, the
trust gain $V_{\rm posterior}^{(n+1)} / V_{\rm prior}^{(1)}$
after $n$ tests necessarily has a range of possible values that grows
geometrically with $n$. In certain cases, a drastic difference
can be obtained by a change of $c_{\rm novel}$: for instance, if instead
of attributing $c_{\rm novel} =100$ to the sixth OFC test, we put 
$c_{\rm novel}=10$ (resp. $1$)
while keeping $p/q=0.1$, $F^{(6)}$ is changed from $4 \cdot 10^{-4}$ 
to $4 \cdot 10^{-3}$
(resp. $0.47$). The trust gain then becomes $V_{\rm posterior}^{(7)} 
/ V_{\rm prior}^{(1)} = 0.07$
(resp. $\simeq 9$). For the sixth OFC test,
$c_{\rm novel}=1$ is arguably unrealistic, given the importance of 
faults in seismology.
The two possible choices $c_{\rm novel} =100$  and $c_{\rm novel} 
=10$ then give
similar conclusions on the invalidation of the OFC model.
In our examples, $V_{\rm posterior}^{(n+1)} / V_{\rm prior}^{(1)}$
provides a qualitatively robust measure of the gain in trust after $n$ steps;
this robustness has been built-in by imposing a coarse-grained
quality to $p/q$ and $c_{\rm novel}$.

\vskip 0.5cm
\noindent
{\bf Summary}

\noindent
The validation of numerical simulations continues to become more
important as computational power grows and the complexity of modeled
systems increases, and as increasingly important decisions are influenced
by computational models. We have proposed an iterative, constructive
approach to validation using quantitative measures and expert knowledge
to assess the relative state of validation of a model instantiated in a
computer code.  In this approach, the
increase/decrease in validation is mediated through a function that
incorporates the results of the model vis-\`a-vis the experiment
together with a measure of the impact of that experiment on the
validation process. While this function is not uniquely specified, it is
not arbitrary: certain asymptotic trends, consistent with heuristically
plausible behavior, must be observed. In five fundamentally different
examples, we have illustrated how this approach might apply to a
validation process for physics or engineering models. We believe that
the multiplicative decomposition of trust gains or losses (given in Eq.
\ref{mglerdffd}), using a suitable functional prescription (such as Eq.
\ref{mgler222}), provides a reasoned and principled description of the
key elements ---and fundamental limitations--- of validation. It should
be equally applicable to biological and social sciences, especially
since it is built upon the decision-making processes of the latter. 
We believe that our procedure transforms
the paralyzing criticisms in Popper's style that 
``we cannot validate, we can only invalidate'' \cite{Oreskes},
espoused for instance by Konikov and Bredehoeft
in the context of groundwater models  \cite{Konikov},
into a practical constructive algorithm which 
addresses specifically both problems of distinguishing between competing
models and transforming the vicious circle of lack of
suitable data into a virtuous circle quantifying the evolving
trust of a model based on the novelty and relevance of new data and 
the quality of fits.

\vskip 1cm
\noindent
{\bf Acknowledgments}:
This work was supported by the LDRD 20030037DR project ``Physics-Based
Analysis of Dynamic Experimentation and Simulation.'' We acknowledge
stimulating interactions and discussions with the other members of the
project, including Bob Benjamin, Mike Cannon, Karen Fisher, Andy Fraser,
Sanjay Kumar, Kathy Prestridge, Bill Rider, Chris Tomkins, Peter
Vorobieff, and Cindy Zoldi. We are particularly appreciative to have
Timothy G. Trucano as our reviewer, who provided an extremely insightful
and helpful report, with many thoughtful comments and suggestions. As
authors, we count ourselves very fortunate to have had such a strong
audience to scrutinize and improve our contribution.

\pagebreak

%FIGURE 1

\clearpage

%FIGURE 2
\begin{figure}
\includegraphics[width=14cm]{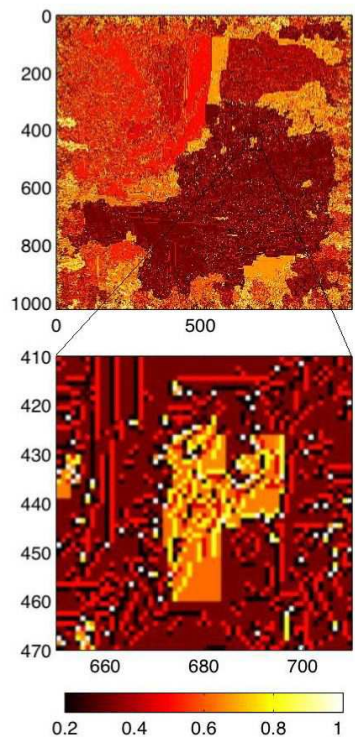}
\caption{(color online)  Map of the ``stress'' field generated by the OFC model
immediately after a large avalanche (main shock)
at two magnifications. The upper
panel shows the whole grid of size $L=1024$ and the lower plot represents
a subset of the grid delineated by the square in the upper plot. Adapted
from \cite{HHS}.
}
\label{OFC}
\end{figure}

\clearpage

%FIGURE 3
\begin{figure}
\includegraphics[width=14cm]{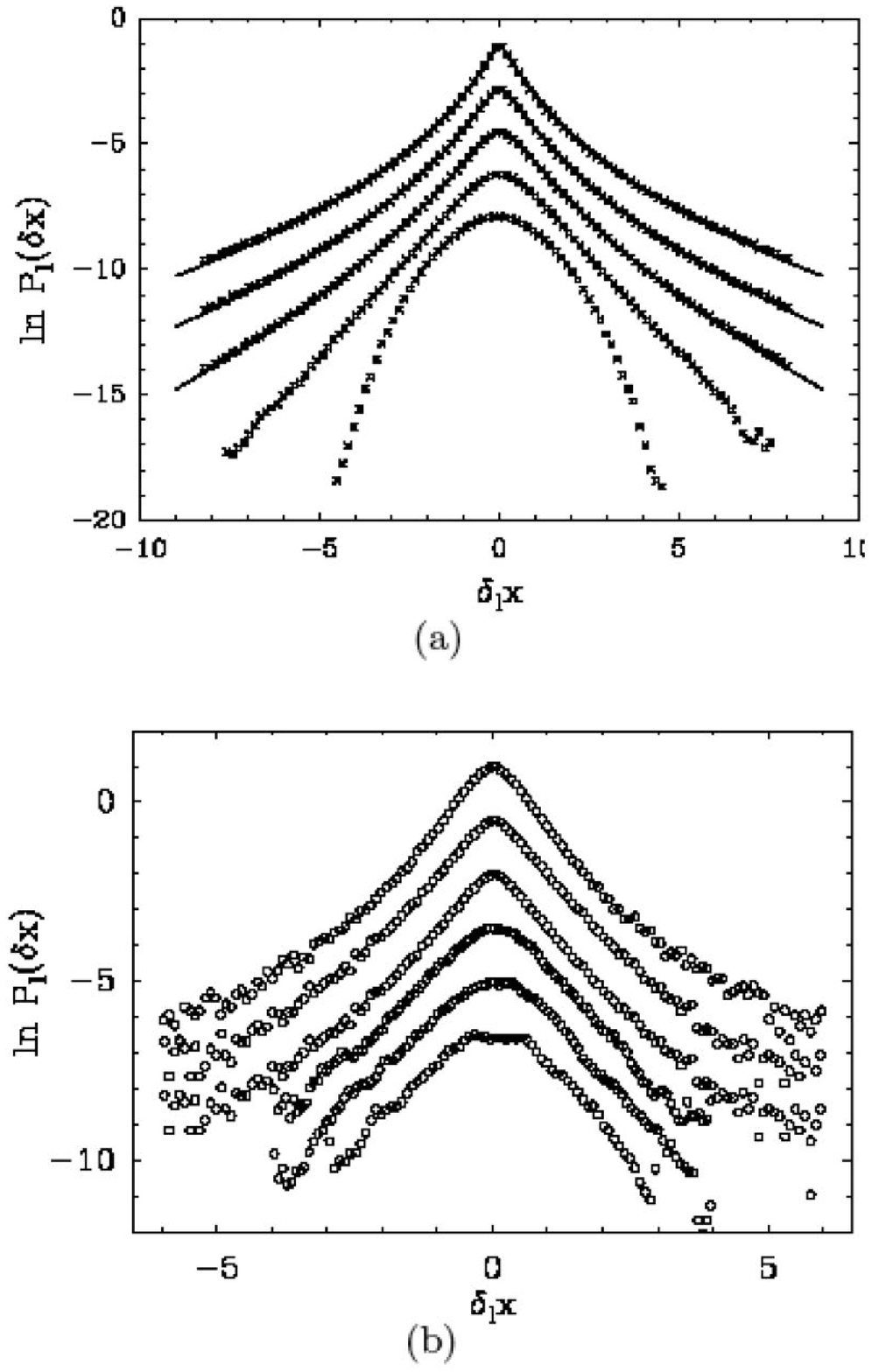}
\caption{Continuous deformation of the PDF of increments
across scales. (a) MRW Model. Standardized PDF's (in logarithmic
scale) of the MRW increments for 5 different time scales (from top to bottom),
$\lambda = 16; 128; 2048; 8192; 32768$. One can observe the continuous deformation
and the appearance of fat tails when going from large
to fine scales. 
(b) S\&P500 future. Standardized PDF's of the returns at scales
(from top to bottom) $\lambda = 10, 40, 160$ min, $1$ day, $1$ week and one
month. As in panel (a), the scale is logarithmic and plots have
been arbitrarily shifted along vertical axis for clarity.
Adapted from \cite{M_etal}.
}
\label{MRW}
\end{figure}

\clearpage

%FIGURE 4
\begin{figure}
\includegraphics[width=14cm]{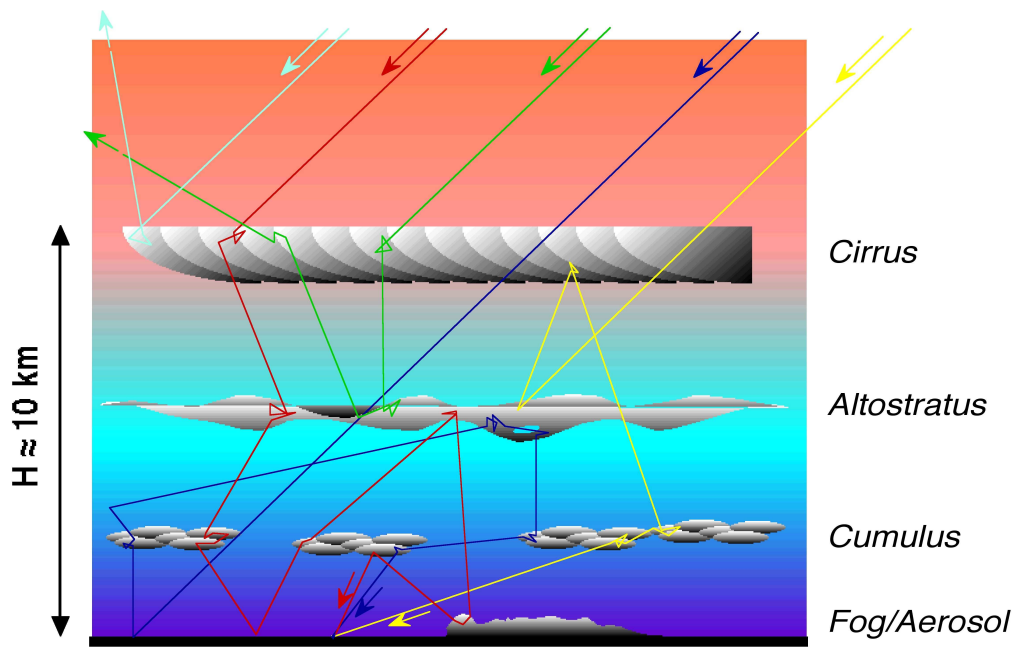}
\caption{Schematic representation of the anomalous diffusion model of solar photon
transport at non-absorbing wavelengths in the cloudy atmosphere. In this model,
solar beams follow convoluted
L\'evy walks, which are characterized by frequent
small steps (inside clouds) and occasional large jumps (between
clouds or between clouds and the surface). The partition between small
and large jumps is controlled by the L\'evy index $\alpha$ (the PDF of the 
jump sizes $\ell$ has a tail decaying as a power law $\sim 1/\ell^{1+\alpha}$
with $1<\alpha<2$).
}
\label{anomalous_diffusion}
\end{figure}

\clearpage

%FIGURE 5
\begin{figure}
\includegraphics[width=14cm]{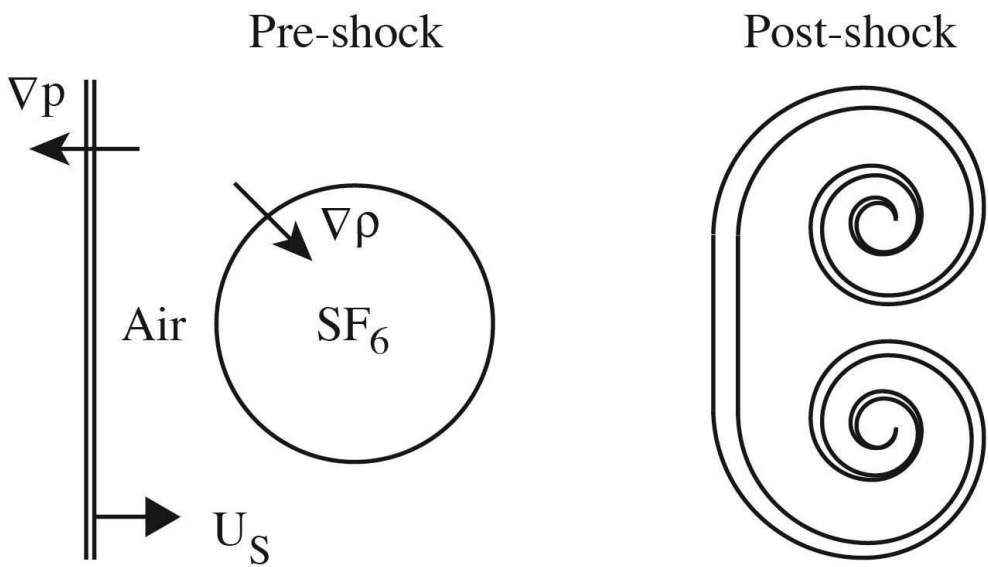}
\caption{Schematic of the interactions between weakly shocked
(Mach number $M=1.2$) light gas (air) and a column of dense gas (SF$_6$ ).
The Richtmyer-Meshkov instability occurs from the mismatch
between the pressure gradient (at the shock front) and the density
gradient (between the light and dense gases), which acts as a source of
baroclinic vorticity. The column of dense gas ``rolls up'' into a
double-spiral form under the action of the evolving vorticity.
}
\label{ShockTubeFig_Kamm_new}
\end{figure}


\begin{thebibliography}{999}
\bibitem{modeldef} By model, we understand an abstract conceptual construction
based on axioms and logical relations developed to extract logical 
propositions and predictions.
\bibitem{footnoteconverg} It is conceivable that a new and radically different
observation/experiment may arise and challenge the built-up trust in a model;
such a scenario exemplifies how any notion of validation ``convergence'' is
inherently local.
\bibitem{satisficing} In economics, ``satisficing'' is a behavior which
attempts to achieve at least some minimum level of a particular
variable, but which does not strive to achieve its maximum possible
value. The word satisfice was introduced by H.
Simon in his theory of bounded rationality.
\end{thebibliography}
\end{document}